\documentclass[12pt,preprint]{aastex}

\usepackage{bm}
\usepackage{xcolor}
\colorlet{myc1}{green!20!red!80!}

\newcommand{\uas}{$\mu$as}

\newcommand{\eb}{\begin{equation}}
\newcommand{\ee}{\end{equation}}

\shorttitle{Radio-optical reference frame objects}
\shortauthors{Makarov et al.}
\begin{document}

\title{The precious set of radio-optical reference frame objects in the light of Gaia DR2 data} 
\author{Valeri V. Makarov, Ciprian T. Berghea, Julien Frouard, Alan Fey}
\affil{United States Naval Observatory, 3450 Massachusetts Ave. NW, Washington, DC 20392-5420, USA}
\author{Henrique R. Schmitt}
\affil{Naval Research Laboratory, 4555 Overlook Ave. SW, Washington, DC 20375, USA}
\email{valeri.makarov@navy.mil}

\begin{abstract}
We investigate a sample of 3413 {\it International Celestial Reference Frame}
(ICRF3) extragalactic radio-loud sources with accurate positions determined by VLBI in the
S/X band, mostly active
galactic nuclei (AGN) and quasars,
which are cross-matched with optical sources in the second Gaia data release (Gaia DR2). The main goal of this study is to
determine a core sample of astrometric objects that define the mutual orientation of the two fundamental
reference frames, the Gaia (optical) and the ICRF3 (radio) frames. The distribution of 
normalized offsets between the VLBI
sources and their optical counterparts is non-Rayleigh, with a deficit around the modal value and a tail extending beyond the 3$\sigma$ confidence level. A few filters are applied to the sample in order to discard double cross-matches,
confusion sources, and Gaia astrometric solutions of doubtful quality. {\it Panoramic Survey Telescope and 
Rapid Response System}
(Pan-STARRS) and {\it Dark Energy Survey} (DES) stacked 
multicolor images are used to further deselect objects that are less suitable for precision astrometry,
such as extended galaxies, double and multiple sources, and obvious misidentifications. 
After this cleaning, 2643 quasars remain, of which 20\% still have normalized offset magnitudes exceeding 3, or
a 99\% confidence level. We publish a list of 2119 radio-loud quasars of prime astrometric quality. The observed dependence of binned median
offset on redshift shows the expected decline at small redshifts, but also an unexpected rise at $z\sim 1.6$, which may be attributed
to the emergence of the C IV emission line in the Gaia's $G$ band. The Gaia DR2 parallax zero-point is found to 
be color-dependent, suggesting an uncorrected instrumental calibration effect.

\end{abstract}


\section{Introduction}
\label{Introduction}
The second release of the Gaia mission data \citep[Gaia DR2,][]{pru,bro} made huge strides both in astrometric accuracy and in the number
of measured objects compared with the first release in 2016. As explained in \citet{lin}, the reference system of the Gaia DR2 astrometric solution,
which defines the orientation of the coordinate triad in space and the rigid spin of the entire ensemble of 
1.7 billion objects, is not independent. 
The coordinates of Gaia sources were adjusted to the International Celestial Reference Frame (ICRF) using a preliminary solution
for the version 3. The proper motions were also adjusted through a
3D rotation using a much greater sample of mid-infrared identified quasars and AGNs from the MIRAGN catalog \citep{sec}. In the latter case, the
constraint is derived from the prior information that quasars should have vanishingly small proper motions due to their large
distances from the observer, except for the small effect of {\it secular aberration} \citep[e.g.,][]{kop}. The former tie, on the other
hand, requires a much more restricted sample of special objects, which we call Radio-Optical Reference Frame (RORF) objects. These should
be radio-loud, unresolved AGNs of supreme astrometric quality, amenable to VLBI position measurements at the microarcsecond level.
At the same time, RORF objects must have sufficiently optically bright counterparts to be observed by Gaia. 

Radio-loud quasars and AGNs are not ideal objects for optical astrometry \citep{mak}. The nearby AGNs are often associated with their
elliptical or spiral host galaxies. The extended substrate image perturbs Gaia astrometry at the centroid fitting level, as the
latter procedure was designed for unresolved, point-like sources in Gaia DR1 and DR2 \citep{fab}. According to the ``fundamental plane"
relation proposed by \citet{ham}, the magnitude
difference in $V$ between the host galaxy and the nucleus is larger for luminous nuclei. The statistical equality is achieved
at $M_V({\rm nuc})=-22.8$, but already at $M_V({\rm nuc})=-25.7$, the host galaxy is typically 2 mag fainter than the nucleus. This result
was based on observations of 70 AGNs with $0.06 \leq z \leq 0.46$. Most of the radio-loud AGNs investigated here have $z>0.5$ and
very luminous nuclei, so the host galaxy perturbation should be much diminished for the bulk of the sample. We also apply additional
vetting (Section \ref{clean.sec}) based on high-quality composite images from {\it Panoramic Survey Telescope and Rapid Response System} (Pan-STARRS) and {\it Dark Energy Survey} (DES) to remove AGNs projected against conspicuous
galaxies. This also helps to get rid of close double and multiple sources, possible lenses, and confusion sources. 

The existence of large differences in the radio-optical positions of reference AGNs of $\sim10$ mas was suspected by \citet{zac}, but the
relatively modest accuracy of the 
dedicated ground-based CCD observations with the 0.9 m telescope could not provide a confident detection. 
Up until the advent of Gaia data, the bulk of optical counterparts of RORF objects were considered to provide high-quality absolute reference.
\citet{ber} used the VLBI-measured
radio positions from the Optical Characteristics of Astrometric Radio Sources (OCARS) compilation \citep{mal,tit,mal16} as hard constraints for a global astrometric
adjustment of the Pan-STARRS catalog and investigated the deviant cases in the process. They determined that a non-negligible fraction
of the VLBI sources ($\sim10$\%) have mismatching optical positions in Pan-STARRS beyond the statistical expectation. A list of sources
was published that should not be used as RORF objects because they exhibit obvious signs of perturbations
on the high-resolution Pan-STARRS maps. The relatively low level of astrometric precision in Pan-STARRS (50--70 mas per coordinate)
did not allow
the authors to study the bulk of cases apart from the gross discrepancies.

The two data releases of Gaia allowed us to look at the problem through the magnifying glass of space-grade astrometry. Based on much more accurate
DR1 data, \citet{pet17} concluded that 6\% of sources with accurate VLBI positions have significant differences at
99\% confidence level, while \citet{mak16} estimated that more than 4\% of the smaller ICRF2 sample  \citep{fey} differ in their radio and optical positions by more than $12\sigma$,
after removing as many contaminants as possible.

In this paper, we review the problem of radio-optical offsets in the light of Gaia DR2 data. Our goal is not to validate
the results of the DR2, as was done by \citet{mig18}. We compute and analyze the absolute and relative position differences
``Gaia $-$ VLBI" in Section \ref{off.sec} and confirm that these differences are even more common and pronounced than what the
previous studies have revealed. After a few types of cleaning and filtering are applied (Section \ref{clean.sec}), the final selection of best quality RORF objects is discussed and
presented in Section \ref{out.sec}. We investigate a
possible relation to the deviation of Gaia parallaxes from the expected zero in Section \ref{par.sec}.
Possible correlation of radio-optical offsets with redshift is addressed in Section \ref{z.sec}, and conclusions are drawn in Section \ref{conc.sec}.

\section{Radio-optical offsets}
\label{off.sec}
\begin{figure}[htbp]
  \centering
  \includegraphics[angle=0,width=0.7\textwidth]{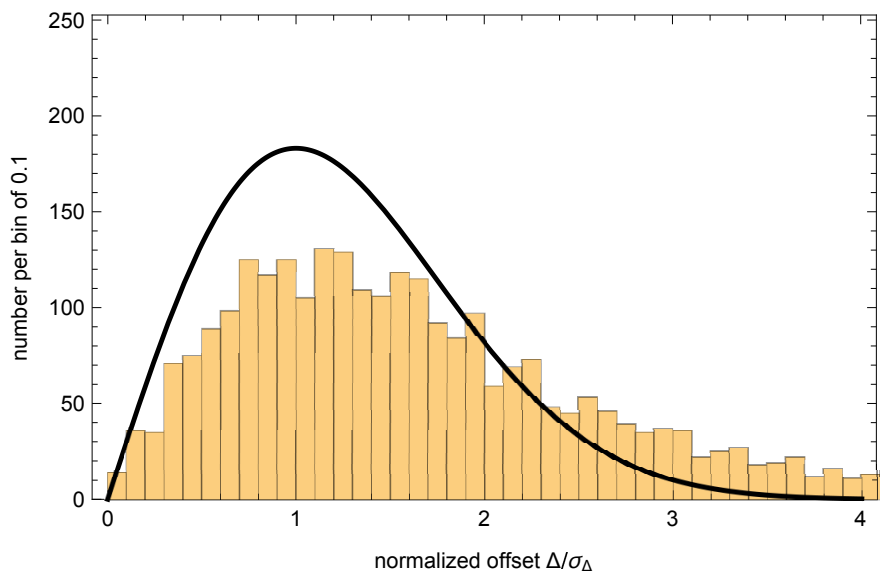}
\caption{Histogram of normalized Gaia--VLBI offsets $u$ (Eq. \ref{u.eq}) for a sample of 3020 astrometric quality RORF objects, after preliminary  cleaning and filtering procedures have been applied. 
The graph is limited to $u < 4$ mas but
the sample distribution extends to above 4000. The black curve shows the scaled Rayleigh distribution with a scaling
parameter $\sigma=1$, which is expected to fit the sample distribution if the coordinate errors are Gaussian-distributed
with the given formal errors.\label{hist.fig}}
\end{figure}

We selected all Gaia DR2 objects within 1$\arcsec$ of each ICRF3\footnote{The S/X ICRF3 catalog is available for download at \url{http://hpiers.obspm.fr/webiers/newwww/icrf/}.} 
source, using the VLBI positions listed in ICRF3. The search radius corresponds to the upper
limit of genuine positional differences found in the previous analysis \citep{mak16}.
This search resulted in $3413$ 
tentative matches, with some of the radio sources associated with more than one Gaia source. After cross-matching this list with
the OCARS compilation\footnote{\url{http://www.gaoran.ru/english/as/ac\_vlbi/ocars.txt}.}
(August 2018 version), we discard all objects labeled as galaxies (G) and BL Lacertae-type (BL), which are known to be problematic
sources for optical astrometry because of the host galaxy contribution in their images. After also cleaning of obvious misses and confusion
sources, $3020$ tentative matches remain.
Following the steps of the previous study based on Gaia DR1 \citep{mak16}, we compute the following quantities for each match: $d_1$ and $d_2$ are the
tangential angular coordinate differences in the sense Gaia $-$ VLBI in right ascension (times cosine
declination) and declination, respectively; $u$ is the normalized total
offset
\eb
u=\Delta/\sigma_\Delta=\sqrt{d_1^2+d_2^2}/\sigma_\Delta. \label{u.eq}
\ee
In the calculation of the formal standard deviation of the absolute offset, $\sigma_\Delta$, the complete $2\times 2$ block of the covariance
matrix was used, as described in \citet{mig16} and \citet{mak16}. This method fully takes into account the covariances 
of coordinate uncertainties in both VLBI and Gaia measurements. The histogram of normalized offsets in Fig. \ref{hist.fig} peaks approximately at the maximum probability density of the expected (for a normal
distribution of coordinate uncertainties) Rayleigh distribution with a scaling parameter of 1. A small shift of the
histogram peak to higher values of $u$ is probably caused by the formal errors of Gaia, which are known to be underestimated \citep{lin}.
The most conspicuous feature is a deficit of values around the mode, which is caused by a long and shallow tail stretching to large offsets. The coordinate differences are definitely not Gaussian, and a significant
fraction of offsets is much greater than what the estimated astrometric precision suggests. A similar behavior, albeit at a much coarser precision,
was seen for the Gaia DR1 positions.

\section{Toward a clean set of RORF sources}
\subsection{Independent vetting}
\label{clean.sec}
\begin{figure}[htbp]
  \centering
  \includegraphics[angle=0,height=0.4\textwidth]{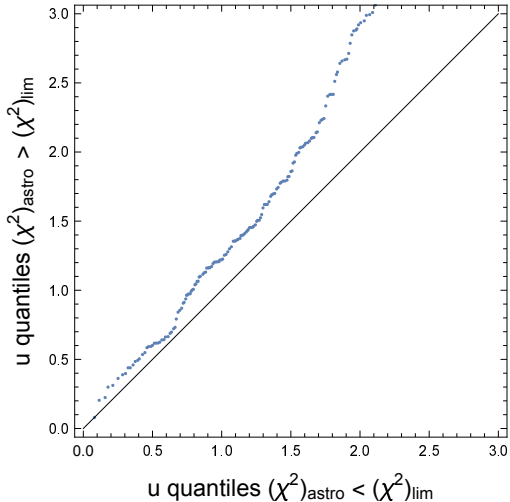}
\caption{Quantile plots of normalized residuals $u$ for RORF sources with high values of reduced $\chi^2$ statistics on astrometric
residuals in Gaia DR2
against the quantiles for sources passing this filter. The straight line represents the locus of quantile values
when the two distributions are identical. \label{qq.fig}}
\end{figure}

\textcolor{myc1}{}As our second step toward a clean sample of RORF objects, we investigated some of the astrometric and photometric criteria suggested in the literature,
aimed at removing perturbed and low-quality solutions in Gaia DR2. A large number of Gaia DR2 astrometric solutions is perturbed
by various factors, including calibration issues, double sources, and crowding contamination. These solutions may not be reliable, in which case they 
should not be used for this analysis. The filter described by \citet[][their Eq. 1]{are}
uses the reduced $\chi^2$ statistic of the residuals and sets up a magnitude-dependent threshold on the excess scatter. Applying this filter
to our sample of 3020 cross-matched Gaia-OCARS sources removes 249 of them, i.e., 8\%. 
To estimate the effectiveness of this filter, we compare the distributions of the normalized offsets $u$ for the sets of discarded sources 
and the 2771 objects that passed the $(\chi^2_{\rm astro})$ filter.   Fig. \ref{qq.fig} shows the results of this comparison represented 
as a ``Q-Q plot", where quantiles of the empirical distribution of $u$ values of the vetted sources (ordinate) are mapped against 
the same quantiles for the 2771 accepted objects (abscissa). The straight line of unit slope shows where the quantile sequence should lie
in case the two independent distributions are identical. The quantile values systematically lie above this line.
This means that the discarded objects have systematically larger values of $u$ at any quantile than the accepted objects.
Thus, we find this quality criterion efficient for cleaning the sample of RORF objects.

The other major quality filter suggested in the literature is that of photometric nature \citep[Eq. 2 in][]{are}, which is aimed at detecting solutions with a significant
impact of extraneous signal using the excess color factor. This filter would remove a hefty 1270 objects, or 46\% of the sample. Analysis of
the position offsets revealed that the filter was not efficient in reducing the scatter or removing extreme outliers. A Q-Q plot
similar to Fig. \ref{qq.fig}, not reproduced here for brevity, shows that quantile values for sources with a $G$-band
excess, referenced to their counterparts for sources without this excess, lie close to the line of unit slope, or even
below it. Removing objects with the photometric excess would not make the distribution of offsets tighter. The inefficiency of this criterion
may be related to the fact that quasars have spectral energy distributions significantly different from field stars. We decided to not apply
this filter to our sample of radio-loud quasars, but to clean it further using other methods independent of the Gaia data. 

We further find that four ICRF3 sources are each cross-matched to two different Gaia DR2 entries. This is not surprising, because the angular
resolution of Gaia measurements is better than $1\arcsec$, which is our search radius. The optical counterparts may be optical pairs
or genuine double AGNs. To avoid possible astrometric errors caused by closely separated
images, we remove all eight cross-matched counterparts from the list.

Following the method first proposed in \citet{mak16}, we collected a large number of high-quality images of RORF objects and visually inspected
all of them. Some of the radio-loud AGNs in our sample reside in the cores of luminous galaxies. Astrometric solutions for extended
objects are perturbed, which possibly explains why the frequency of outliers is higher at small redshifts (Fig. \ref{quant.fig}). 
OCARS provides specific morphological classification of the optical counterpart, which we already used to remove all objects flagged as
``galaxies". The major source of images came from the collection of colored stacked 
(Pan-STARRS) available through the Mikulski Archive for Space Telescopes (MAST) 
provided by the STScI. Pan-STARRS includes a multipassband panoramic survey of the northern 3/4 of the sky \citep[above Dec$\simeq -30\degr$,][]{cha}
so only three quarters of the sample can be reviewed this way. For the southern sky, we used the recently published access facility to
stacked (DES) images \citep{mor}. The quality of both Pan-STARRS and DES images is superior to the previous 
surveys, routinely reaching better than $1\arcsec$ resolution. 

\begin{figure}[htbp]
  \centering
  \includegraphics[angle=0,height=0.32\textwidth]{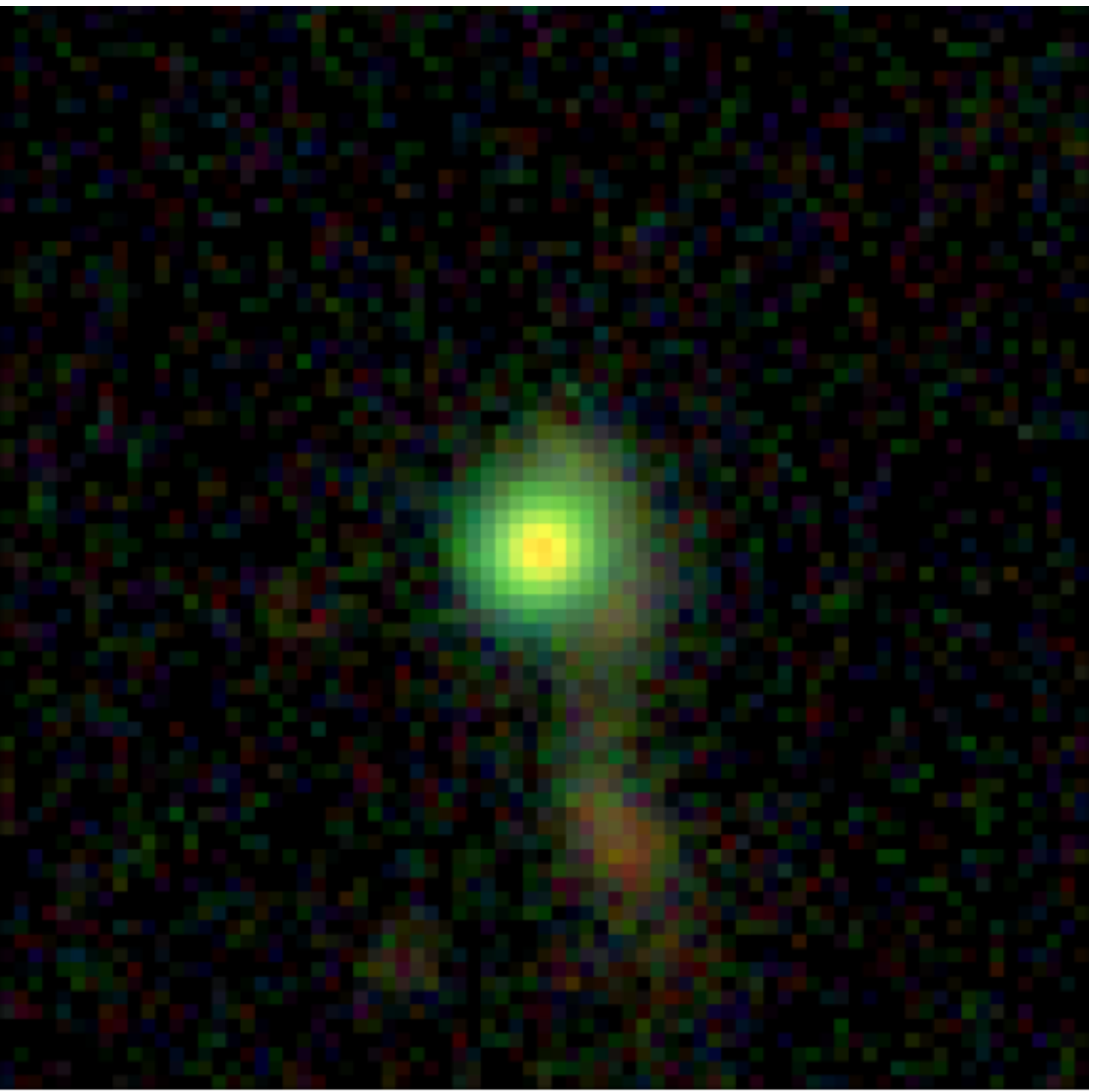}
  \includegraphics[angle=0,height=0.32\textwidth]{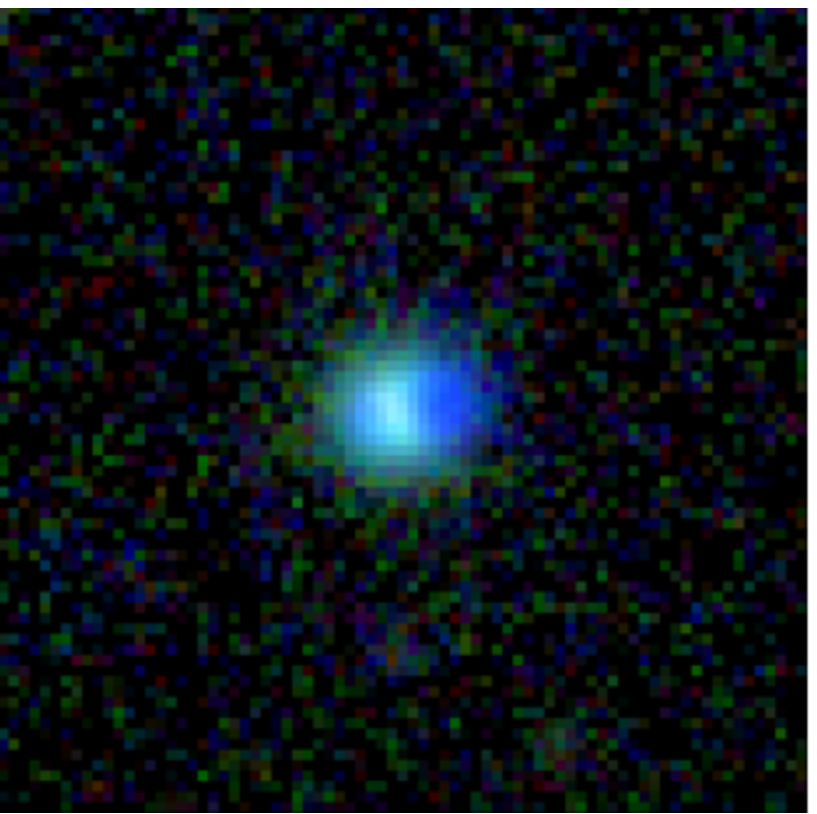}
  \includegraphics[angle=0,height=0.32\textwidth]{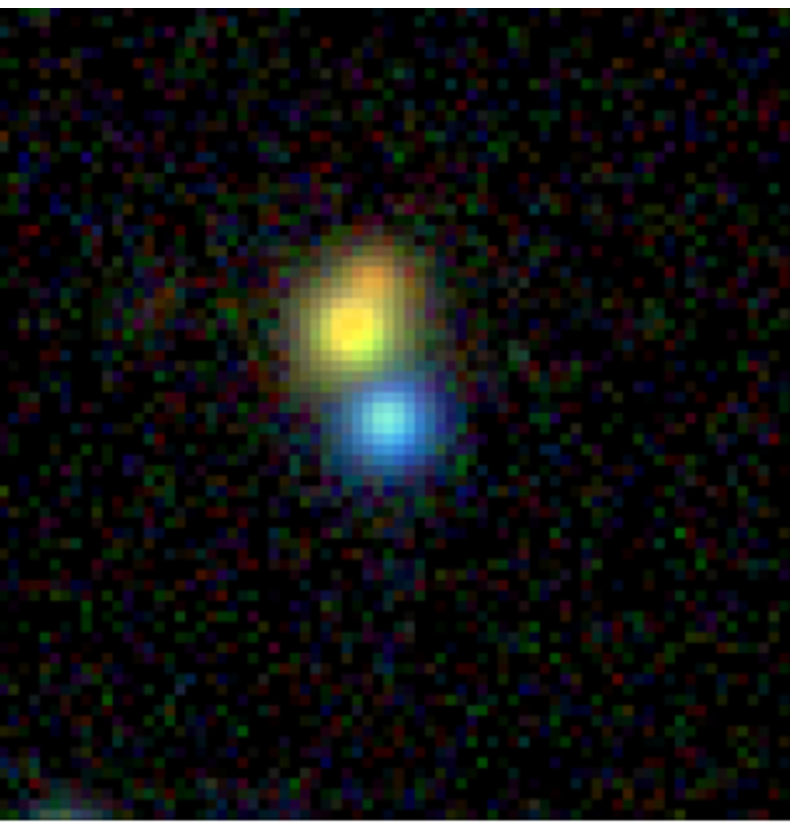}
\caption{DES images of three representative types of optical counterparts of VLBI-measured radio-loud AGN, which are deemed not
suitable for a RORF tie. Each image
is $20\arcsec$ on a side, with north up and east to the left. The matched radio source is
exactly at the center of each image. Left: PKS $0521-403$, an interacting galaxy.  Middle: IVS B$0307-362$, possibly double
or microlensed. Right: IVS B$2203-617$, a close multiple object. \label{map.fig}}
\end{figure}

We performed a blind inspection of collected images without referencing the position offsets. The images were separated into four groups, based on
the visual inspection. One group was comprised of objects where the presence of a host galaxy was obvious to the eye. Tightly spaced double or
multiple sources were separated into another group. We also found a smaller number of objects where no optical counterpart was visible at the
VLBI location, but a faint optical source (likely, a chance field star) was present at separations $s\ga 0\farcs 5$. Finally, the largest
group included sources that did not reveal any obvious problems and looked point-like. Fig. \ref{map.fig} shows a few examples of DES images,
which indicate problematic cases for optical astrometry. Any deviation from a nominal star-like image can significantly perturb the Gaia-determined
photocenter, because a single template line spread function
was applied for each CCD and each gate \citep{fab} in the low-level pipeline processing. These perturbations are not always captured by the
$\chi^2$-based residual statistic, apparently.

To estimate the effectiveness of the blind image vetting,
we compare the distributions of the normalized offsets $u$ for the sets of extended sources and double or multiple sources using the empirical distribution
of star-like sources as reference. The corresponding Q-Q plots indicate that the vetting of remaining extended optical
images is moderately successful, as the quantile values for them are systematically higher than the quantiles for
point-like sources. On the other hand, the vetting of resolved double sources is not efficient, removing only several
extreme outliers at $u>4$. It appears that double images at separations greater than $1\arcsec$ do not much perturb
Gaia astrometric solutions. However, we decided to remove all identified double, extended, and misplaced images from the
sample in view of their relatively small number. It should be noted that for ICRF3 sources in some areas south of Dec$= -30\degr$
without DES coverage, no images of comparable quality are available.   

\subsection{The precious set}
\label{out.sec}
The final ``clean" sample counts 2643 RORF objects.
Thus, the cleaning removed 23\% of the starting selection. The histogram of normalized offsets $u$ in Fig. \ref{hist.fig} 
is still quite far from the expected Rayleigh$(1)$ distribution. The pronounced deficit of values in the core is caused by the excess of statistically large
position offsets. This result can be expressed in terms of the statistical survival function, which estimates that
99\% of a Rayleigh-distributed sample should have values less than 3. In fact, 20\% of the cleaned sample (524 sources)
have normalized position differences in excess of 3. The rate of statistically significant offsets is dramatically
larger than the previous estimate based on Gaia DR1. The reason for this is probably the much improved accuracy of Gaia DR2
positions compared with DR1. We can more clearly see now that the bulk of radio-loud quasars have their optical photocenters
displaced at the submilliarcsecond level.

Still, these objects straddling the radio and optical domains, provide the best chance to establish an accurate reference
frame tie between Gaia and VLBI, as far as accurate positions are concerned. We therefore suggest an additional empirical cut
of the sample at $u=3$ to generate the cleanest set of RORF objects, which pass all the currently available quality criteria.
This ``precious set" of 2119 sources is published in its entirety online. Table~1 provides a small cutout of the file.
VLBI coordinates from ICRF3 (columns 2 -- 3) and redshifts (column 10) are combined with astrometric and photometric
information from Gaia DR2 (columns 4 -- 9) followed by position differences (columns 11 -- 13) derived in this paper.

\begin{deluxetable}{lrrrrrrrrrrrl}
\tablecaption{Best quality radio-optical reference frame objects \label{obj.tab}}
\tablewidth{0pt}
\tablehead{
\multicolumn{1}{r}{(1)$^\dag$}  &
\multicolumn{1}{r}{(2)}  &
\multicolumn{1}{r}{(3)}  &
\multicolumn{1}{r}{(4)}  &
\multicolumn{1}{r}{(5)}  &
\multicolumn{1}{r}{(6)}  &
\multicolumn{1}{r}{(7)}  &
\multicolumn{1}{r}{(8)}  &
\multicolumn{1}{r}{(9)}  &
\multicolumn{1}{r}{(10)}  &
\multicolumn{1}{r}{(11)}  &
\multicolumn{1}{r}{(12)}  &
\multicolumn{1}{c}{(13)}   \\
}

\rotate \tabletypesize{\scriptsize} \startdata
$0948+658$ & 148.134135797 & 65.6336714927 & 1066310912103182464 & 0.110 & 0.145 &
   18.034 & 18.295 & 17.526 & nan & $-0.827$ & 0.892 &
   0.763 \\
$1016+635$ & 154.961986900 & 63.3337852583 & 1052151882396770944 & $-0.041$ & 0.161 &
   18.462 & 18.568 & 17.590 & 2.025 & $-0.073$ & $-0.017$ & 0.299
   \\
$0839+687$ & 130.954589771 & 68.5547652916 & 1118095435870099584 & $-0.539$ & 0.625&
   20.010 & 20.089 & 19.571 & nan & $-0.278$ & 0.969 &
   1.806 \\
$0859+681$ & 135.971479324 & 67.9563015749 & 1117118691587459200 & $-0.175$ & 0.195 &
   18.313 & 18.613 & 17.722 & 1.499 & 0.120 & 0.194 & 1.09 \\
$0928+653$ & 143.227406594 & 65.1281379721 & 1067919119657621504 & nan & nan &
   20.996 & 21.107 & 19.888 & nan & $-0.915$ & 0.060 &
   0.770 \\
$0810+646$ & 123.663293069 & 64.5227850264 & 1091852056117623936 & $-0.037$ & 0.046 &
   15.869 & 16.269 & 15.173 & 0.239 & $-0.181$ & $-0.028$ & 0.547
   \\
$0759+641$ & 120.967330777 & 64.0539924397 & 1091417229331516032 & 0.027 & 0.304 &
   18.889 & 19.046 & 18.398 & nan & $-0.014$ & 0.450 &
   1.592 \\
\enddata
\tablenotetext{\dag}{Columns:\newline
(1) IERS designation; (2) ICRF3 right ascension, deg; (3) ICRF3 declination, deg; (4) Gaia unique source
identifier; (5) Gaia parallax, mas; (6) Gaia parallax
error, mas; (7)  G magnitude; (8) BP magnitude; (9) RP magnitude; (10) redshift $z$ from OCARS; (11) RA difference (times $\cos\delta$)
``Gaia$-$ICRF3", mas; (12) Dec difference ``Gaia$-$ICRF3", mas; (13) normalized position difference $u$.}
\end{deluxetable}

\section{The parallax zero-point}
\label{par.sec}
Quasars can be considered zero-parallax objects because of the great distances separating them from the Sun. The parallaxes of the radio-loud
sources from OCARS measured by Gaia provide a method to estimate the so-called zero-point error, which is a constant bias applied to all
objects. Based on a much larger selection of optically identified quasars from \citet{sec}, \citet{lin} estimated the global zero-point at
$-29$ \uas. Individual parallax determination can be affected by blended confusion sources and extended structures of the host galaxies.
Here we use our clean sample (before the $u<3$ cut) to estimate the parallax zero-point for 2465 radio-loud sources.

\begin{figure}[htbp]
  \centering
  \plotone{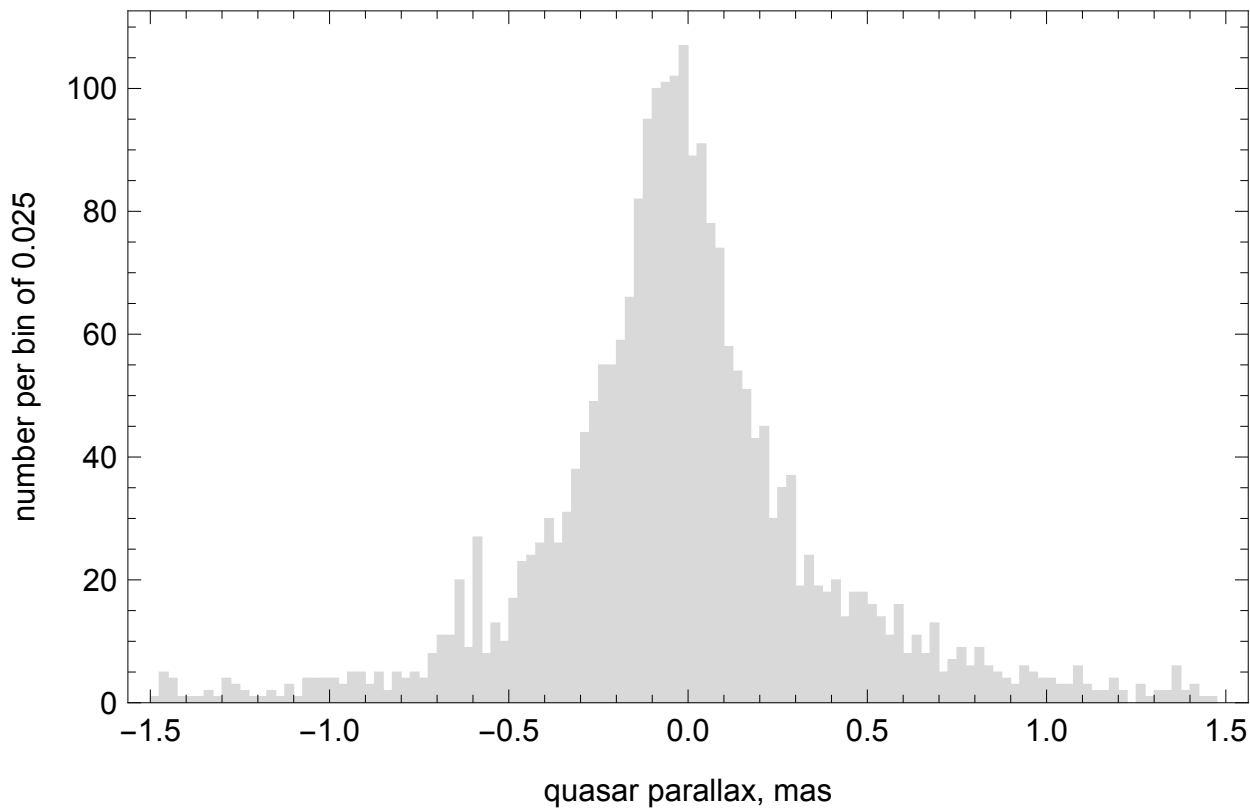}
\caption{Histogram of Gaia DR2 parallaxes for 2465 counterparts of ICRF3 radio-loud quasars. \label{par.fig}}
\end{figure}

\begin{figure}[htbp]
  \centering
  \plottwo{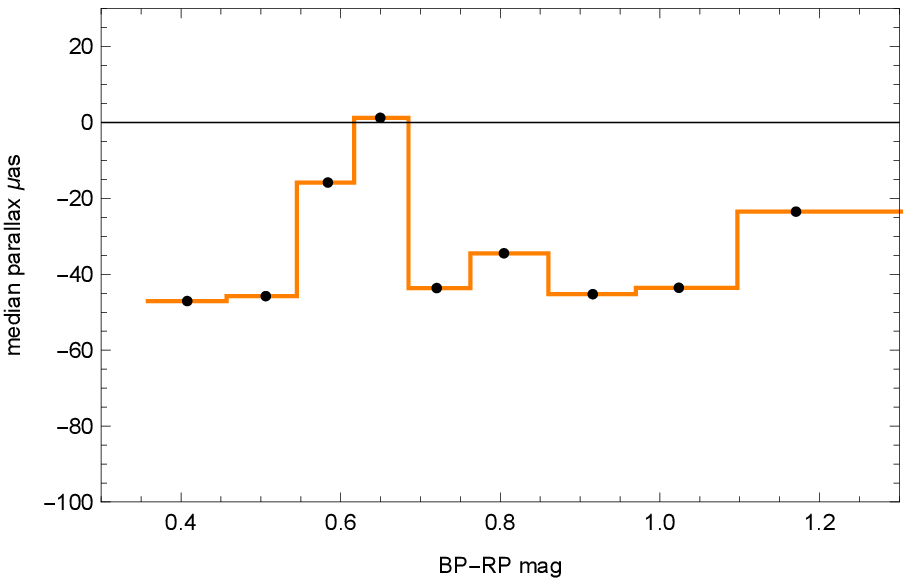}{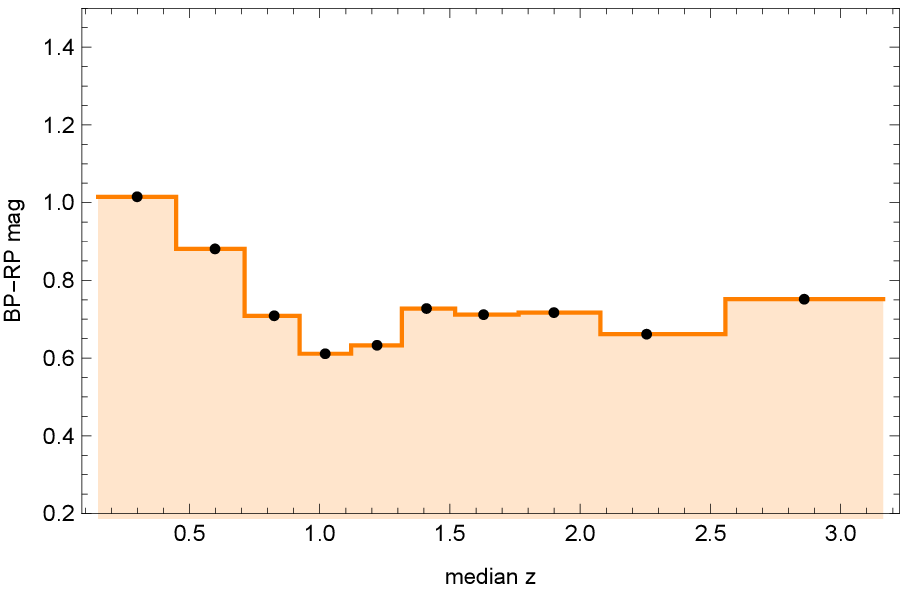}
\caption{Left: binned median parallax of ICRF3 sources versus color BP$-$RP in Gaia DR2. The bin size is 246 objects. The estimated error of the mean parallax for each bin is 16 $\mu$as. Right:
binned median color BP$-$RP  in Gaia DR2 versus redshift $z$. The bin size is 205 objects. \label{cloud.fig}}
\end{figure}

The histogram of measured parallaxes in Fig. \ref{par.fig} has a complex non-Gaussian shape because it is composed of objects from a range of magnitudes, whose
intrinsic astrometric precision also varies over a wide range. The core of the distribution is sharp, and it is obviously shifted
to the negative side. We used two kinds of robust statistics to estimate the ``average" shift. The median of all parallaxes is equal
to $-35$ \uas, and the biweight location\footnote{\url{http://docs.astropy.org/en/stable/api/astropy.stats.biweight\_location.html}.} (with a scaling parameter of 6) is also $-35$ \uas. The closeness of these estimates indicates a stable
result, but they are somewhat larger in magnitude than the previous \citep{lin} estimate $-29$ \uas. Although the zero-point error is
expected to originate from a specific time-dependence of the Gaia basic angle, which is hard to calibrate with the desired accuracy,
the color-dependent calibration term may interfere with this parameter too. The large collection of MIRAGN objects, selected by their
midinfrared colors, may have different optical colors from the radio-loud VLBI AGNs. The latter group shows a large degree of
variability in magnitudes and colors, and a complicated dependence of ``quiescent" magnitudes with redshift $z$ (Berghea et al.,
in preparation). There seem to be two
subpopulations of ICRF3 sources segregated in the color-redshift plane, with the nearby objects (small $z$) being typically redder
and less variable than the more distant ones. This segregation also shows in Fig. \ref{cloud.fig}, right panel, where we used the
nominal BP$-$RP colors from Gaia DR2 and the redshifts of 2051 objects to estimate how the median color depends on redshift. We detect
a sharp transition from predominantly red to much bluer colors at approximately $z=0.65$. The bluest quasars have $z$ around 1, while
the more distant ones at $z>1.4$ do not show much variation in color. 

Fig. \ref{cloud.fig}, left panel, displays the median parallax of 2466 ICRF3 objects (with carefully cleaned and vetted
optical counterparts) for bins of 
sorted BP$-$RP color of equal size (246). We find a complex and unexpected behavior of the parallax zero-point error with the object's color.
Starting below $-40$ \uas\ for the bluest quasars, the parallax increases to zero at BP$-$RP$\approx 0.6$, but it suddenly drops back
to below $-40$ \uas\ at BP$-$RP$\approx 0.7$. This sharp transition hints at an instrumental effect. The average zero-point value for our
sample results from this strong dependence and the distribution of colors, but the latter is sample-specific, so other studies
can indeed arrive at different estimates in the range between 0 and $-50$ \uas. Apparently, there is a calibration issue comparable
in magnitude to the zero-point bias itself.

\section{A cosmological factor in the radio-optical offsets?}
\label{z.sec}
Both Gaia DR1 and DR2 data show a large excess of position differences outside of the expected dispersion due to purely astrometric
errors. Based on the Gaia DR1 astrometric data, \citet{mak16} estimated that more than 4\% of ICRF2 sources with optical counterparts
passing all available quality filters (single, point-like, unperturbed) have position differences outside of the statistical
expectancy. It was suggested that the AGNs in these distant objects are physically ``dislodged", i.e., they are not located in the
optical centers of the host galaxies. This hypothesis will be testable when high-precision epoch astrometry from Gaia becomes available
in the future releases. Some of the AGNs are highly variable in the optical, while the host galaxy is constant. The photocenter of a
variable source blended with an offset constant source displays a coherent  Variability-Induced Motion (VIM) effect \citep[e.g.,][]{makvim}, where the astrometric
displacement is correlated with the light curve. An alternative explanation was proposed by \citet{pet} and citet{pla}, where the observed displacement
is caused by milliarcsecond-scale jets luminous enough in the radio to cause a photocenter shift. With different parts of a relativistic
jet being responsible for the most compact source of emission detected by the VLBI, the optical photocenter can be shifted from
a relatively larger contribution of the accretion disk.

\begin{figure}[htbp]
  \centering
  \plottwo{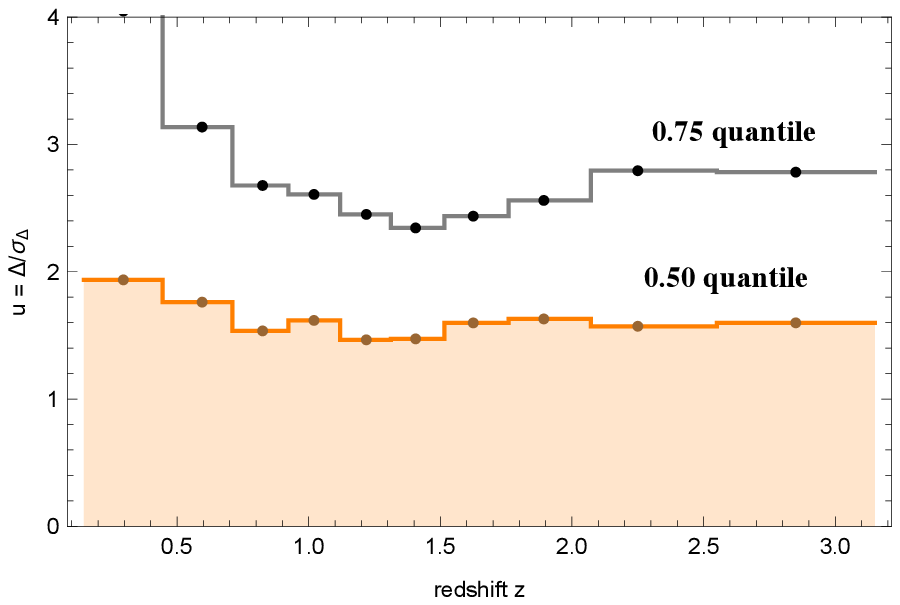}{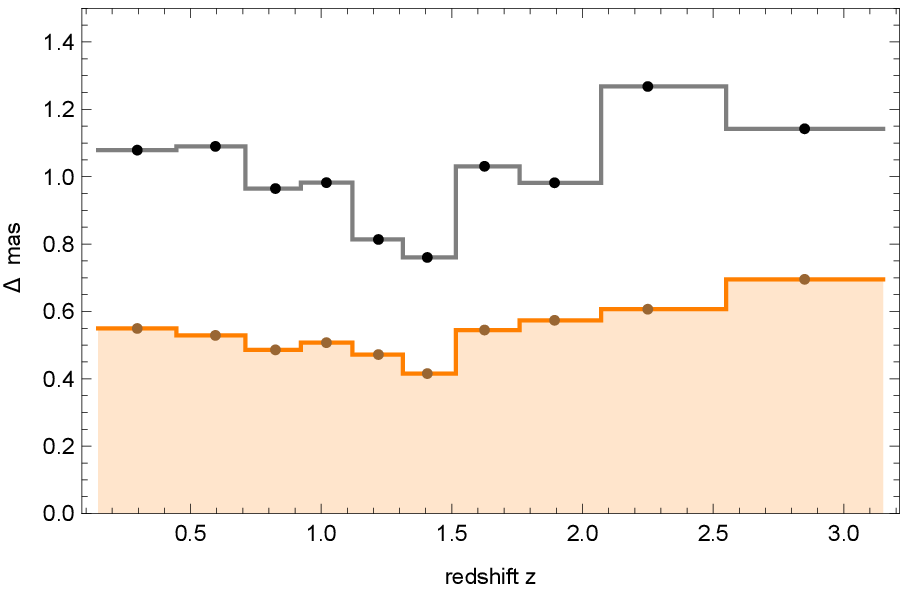}
\caption{Median (0.50 quantile) and 0.75 quantile of  relative offsets $\Delta /\sigma_\Delta$ (left) and absolute offsets 
$\Delta$ in mas (right)
versus redshift $z$ for cross-matched ICRF3 sources in Gaia DR2. The estimated uncertainties of the medians
for each bin are $0.063$ for $u$ and $0.030$ mas for $\Delta$. \label{quant.fig}}
\end{figure}

Both these interpretations can be indirectly tested using the observed relation between the offsets and cosmological redshifts. The OCARS
catalog conveniently includes the redshifts for most of the objects collected from the literature. We are using the redshifts only for large-number statistical
estimation. The starting assumption is that the observed angular displacement depends on the distance to a given AGN, if the typical
physical displacement (or jet) has a certain characteristic size in the object's comoving frame. More distant objects would be
statistically less displaced in the angular measure, but the spacetime is not flat, and a more complicated relation emerges, depending
on the cosmological model of the universe. The relevant parameter is the ``angular diameter distance", which is a nonlinear function of
redshift. Most of the ``standard" models predict that this distance rapidly rises with $z$ at small $z$, peaks at roughly $z=2$, and
becomes either flat or is slowly declining at higher redshifts. This should give a concave curve on a ``offset versus redshift"
diagram. To test this prediction, we performed the following calculation. The vetted
quasars with available redshifts (counting 2074 objects) were sorted by $z$ and divided into 10 equal  bins of 207 objects each. Each
subsample covers a certain range of redshifts. The $0.50$ (median) and $0.75$ quantiles of the normalized offset $u$ were computed for each
bin of redshifts. The results are shown in Fig. \ref{quant.fig}. Both quantile offsets become smaller with increasing $z$ at small
values, as expected, but the decline is rather slow. The offsets reach a minimum at $z\simeq 1.5$, followed by a small
step-up. This increase in offsets is more visible in absolute position differences ($\Delta=\sqrt{d_1^2+d_2^2}$), right plot, than in
normalized offsets $u$, left plot. The jump of median $\Delta$ at
1.5 is from $0.42$ to $0.58$ mas. For $z>1.6$, the dependence becomes rather flat for $u$ and steadily increasing for $\Delta$. These
differences are possibly caused by the distribution of $G$ magnitudes, which become fainter with increasing $z$, making the formal errors
of Gaia rapidly grow.

\begin{figure}[htbp]
  \centering
  \plotone{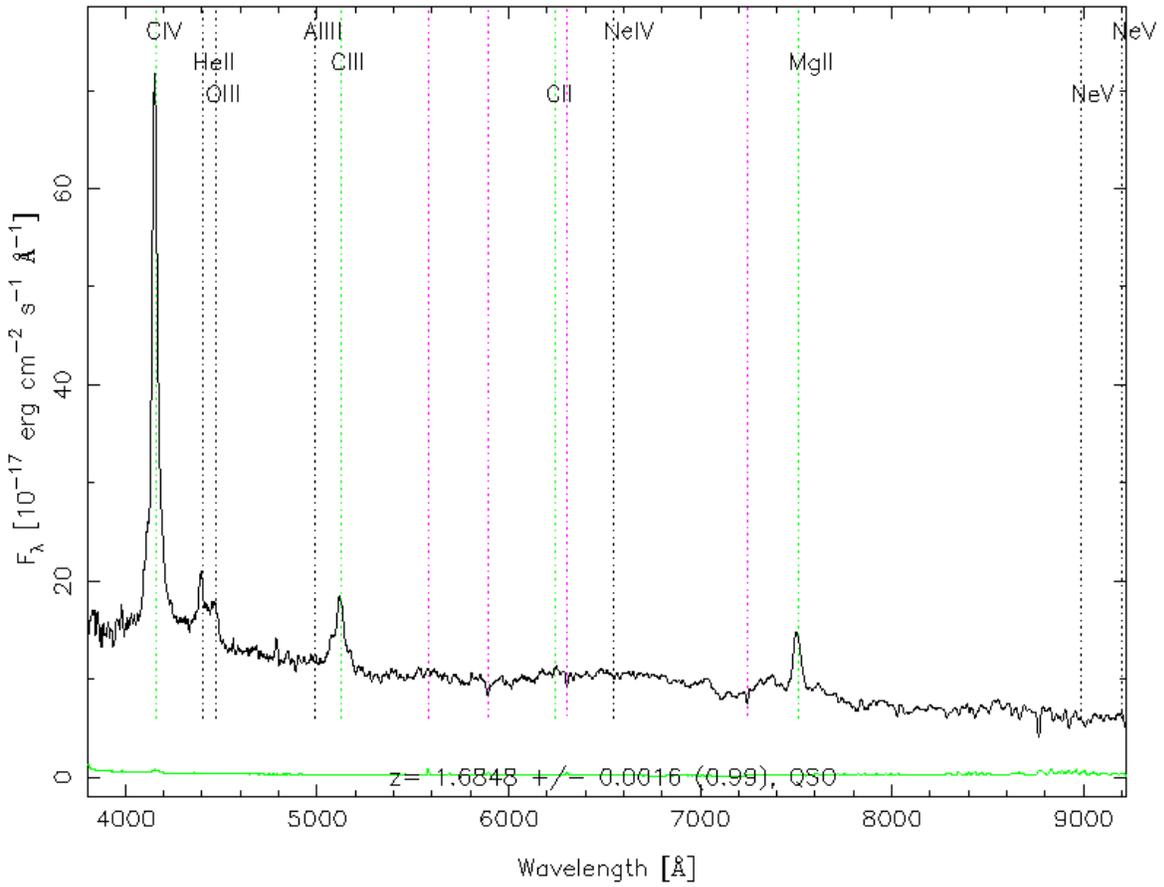}
\caption{SDSS7 spectrum of the ICRF3 QSO $1319+220$ at $z=1.685$. The emission spectrum is dominated by the C IV line, which
appears at the blue edge of the $G$-band sensitivity window for $z > 1.6$. \label{spe.fig}}
\end{figure}

This relation may seem at odds with the predictions of standard models. Of course,
our starting assumption that the physical size of displacement is independent of $z$ may be incorrect. But the experiment indicates
a nonmonotonic factor missing in the model. It is possible that the measured offsets (or their formal errors) are subject to
an unidentified instrumental error in the Gaia data. This error may be magnitude-dependent. The quasars become generally fainter with increasing $z$ at redshifts greater than 1.2, but the dependence is fairly
smooth and no feature is visible around $z=1.5$. An instrumental systematic error can also be color-dependent. \citet{lin} describe
the limitations of the color-dependent calibration in the Gaia DR2. A similar analysis of the median color BP$-$RP versus redshift
is shown in Fig. \ref{cloud.fig}, right panel. It reveals that the nearest radio-loud quasars are red, but the color becomes bluer
by almost 0.4 mag at $z\simeq 1$, where a global minimum is observed. The color becomes slightly redder again by $z\simeq 1.5$, after
which it becomes flat or slightly declining. This complex behavior of color versus $z$ may look unexpected, but photometric analysis
of completely independent Pan-STARRS data (C.T. Berghea et al. 2019, in preparation) shows a similar pattern in the $r_{\rm PS1}-z_{\rm PS1}$
color. The rise at $z\simeq 1.4$
may be caused by the relatively bright emission line Mg II $\lambda 2798$, which shifts from the BP to the RP spectral
window (roughly corresponding to the Sloan Digital Sky Survey --SDSS-- $i$ band) at this redshift. Similar small reddening in median colors have been noted in other photometric investigations
\citep{cro,meu}. For the broad $G$ band, where Gaia astrometric measurements are taken, the C IV emission line may be of
greater importance, emerging within the band at $z\ga1.6$. Fig. \ref{spe.fig} displays the spectrum of IERS $1319+220$, which is one
of the quasars that pass all the quality criteria described in this paper, but are not included in the precious set on account of
elevated offsets ($u>3$). At $z=1.685$, the dominating C IV emission line is well within the $G$-band at its cutoff around 400 nm. If
the source of emission in this line is physically displaced from the source of the radio emission, a measurable offset in positions
occurs. The cross-over of the C IV line into the Gaia astrometric band happens at $z=1.6$, as can be seen from comparison of SDSS spectra
for IERS $1235+196$ at $z=1.533$ and $1232+366$ at $z=1.598$. Other emission line cross-overs of note are for the lines C III at $z=1.1$
and Ly$_\alpha$ at $z=2.3$. For example, the spectrum of the ICRF3 QSO $1337+637$ ($z=2.56$) is dominated by a Ly$_\alpha$ line at
approximately 425 nm. There are also strong C IV and C III lines present within the $G$ band. This object has a large Gaia--ICRF3 offset, which
left it out of the precious set. We do not find signs of astrometric perturbation associated with the emergence of this line in Fig.
\ref{quant.fig}, except for a possible bump in the 0.75 quantile of absolute offsets (right plot).

\section{Conclusions}
\label{conc.sec}
Our previous analysis based on Gaia DR1 and ICRF2 data \citep{mak16} determined that the core distribution of Gaia--VLBI position offsets was
consistent with the expected PDF, but a significant
fraction of the matches had large differences in positions that could not be explained by the estimated random or systematic
errors. The tremendous improvement of astrometric precision achieved in Gaia DR2 for faint objects \citep{lin}, where its level
is already close to the expected 5 yr mission performance \citep{bro}, reveals a greater rate of radio-loud objects with
significant offsets in optical positions. Even after a rigorous cleaning of the sample with astrometric quality filters
and image-based vetting, we arrived at a sample that includes 20\% of objects with normalized differences above 3,
where the expected rate of outliers is 1\%. This implies that a significant fraction of RORF
objects, which can be accurately measured both by VLBI and Gaia, are perturbed by a ``cosmic error" (called
DARN in \citet{zac}). The origin of this perturbation
may be in the nonstellar morphology of the sources. The angular resolution of Gaia is still much lower than that of VLBI, and
the galactic substrates, blended cores of optical emission, gravitational lenses, and jets, can cause the optical photocenters to
shift from the more constrained radio positions. A fraction of RORF objects can also be physically displaced with respect to
their host galaxy centers \citep{ski}. This may be the main explanation for relatively nearby AGNs ($z \la 1$), which
are typically redder with a significant contribution of their host galaxies in the flux. Additionally, 
the Gaia low-level data pipeline has not been tuned to centroid complex or extended
sources. \citet{pet18} used a much larger compilation of VLBI sources (but with less precise positions)
for a similar study. They estimated the rate of statistically significant offsets at 9\%, but this estimate
was obtained after inflating both the VLBI and Gaia formal uncertainties. It is likely that the ICRF3
positions are more precise than the larger collections of VLBI positions, which allows us to avoid modifications
to estimated coordinate variances in this analysis.
A preliminary solution for ICRF3 positions analyzed by \citet{fro} showed a closer agreement with Gaia DR1 than ICRF2 positions.

Once the reality of this perturbation is established, the strategy of using RORF needs to be changed. They are still indispensable
as the main method to tie the two reference frames. Those quasars that have large position offsets, as interesting and enigmatic
they may be from the astrophysical point of view, are practically useless as RORF objects. Their inclusion would do more harm
than good to the effort of bringing together the reference frames. \citet{pet18} also caution against
using RORF sources as absolute reference for proper motions, because photometric variability of the central
emission-line regions may cause the optical photocenter to move mimicking astrometric angular motion.
In this paper, we determined and published a set of 2118
radio-optical quasars that are not perturbed too strongly, and, therefore, can still be used to determine the orientation
of the Gaia coordinate system in space.

The improved precision of the optical data also allows us to look at the possible cosmological relation of the offsets.
The median position offset (and its higher quantiles), both in absolute and relative measure, shows an unexpected feature
at redshift $z\approx 1.5$, where the generally concave and smooth dependence appears to be broken. We suggest
that this peculiarity may be related to the emergence of the C IV emission line at the blue edge
of the $G$-band spectral sensitivity window. We examined multiple optical spectra of the 524 sources
that pass the astrometric quality criteria but still have large optical-radio position offsets.
There seems to be two broad categories of spectra present, namely, relatively nearby objects with weak
or invisible emission lines (e.g., IVS $1204+057$), or QSO with powerful redshifted Mg II ($z>0.45$), C III
($z>1.1$), C IV ($z>1.55$), or
even Ly$_\alpha$ ($z>2.5$) located closer to the blue end of the astrometric band.
The former category may comprise BL Lac-type objects, despite our conscious effort to remove them early in
the analysis, which are known to have larger astrometric offsets from previous publications.
The latter category indicates that the compact sources of radio-emission that form VLBI sources
may be displaced from the broad-line. To confirm or disprove this
behavior, more accurate and internally consistent astrometric observations of RORF objects are needed, especially
at declinations less than $-30\degr$, where ICRF3 is conspicuously sparse.
Future Gaia data releases may help to clarify the matter, although the advance in astrometry is not expected to be drastic
for these mostly faint objects. High-resolution imaging has proven helpful for this analysis, and we have made a first
step with collecting a library of best-quality optical images of these radio-loud sources. A more consistent effort with
space-borne telescopes may be justified in the future, given the importance and relative scarcity of RORF objects.
Finally, a systematic spectroscopic and photometric investigation of ICRF3 sources is in order, especially
in the sky areas outside of the SDSS footprint.

If the reason for the commonly present increased perturbation of radio-loud quasars is based in the astrometric properties
of Gaia, it should also be reflected in the observed parallaxes and proper motions. These data have been used for verification and
validation of Gaia DR2 data \citep{mig18}. We do not find any significant correlation between the error-normalized position
offsets and parallaxes within our precious set of RORF objects. This supports the notion that the large offsets are intrinsic.
The negative bias of quasar parallaxes is quite prominent for this sample, and we find it to be color-dependent in a complex
way, hinting at a chromatic calibration issue in Gaia.

Low-redshift sources do have distinctly greater radio-optical offsets. This can be explained in different ways, including a relatively
larger contribution of extended galaxy images, but the most straightforward interpretation assumes that a certain physical separation
between the most luminous components yields a smaller angular resolution in the observer's frame at greater
distances. A large fraction of radio-loud
quasars display rather extreme variability on the time scales of a month and less, both in the optical and the $X$-band \citep{bar}.
Comparative analysis of simultaneous photometric observations in the optical and radio passbands could perhaps reveal if
the same components of AGN machines are observed.

\section{Acknowledgments}
This work has made use of data from the European Space Agency (ESA)
mission {\it Gaia} (\url{http://www.cosmos.esa.int/gaia}), processed by
the {\it Gaia} Data Processing and Analysis Consortium (DPAC,
\url{http://www.cosmos.esa.int/web/gaia/dpac/consortium}). Funding
for the DPAC has been provided by national institutions, in particular
the institutions participating in the {\it Gaia} Multilateral Agreement. The Pan-STARRS1 Surveys (PS1) have been made possible through contributions of the Institute for Astronomy, the University of Hawaii, the Pan-STARRS Project Office, the Max-Planck Society and its participating institutes, the Max Planck Institute for Astronomy, Heidelberg and the Max Planck Institute for Extraterrestrial Physics, Garching, The Johns Hopkins University, Durham University, the University of Edinburgh, Queen's University Belfast, the Harvard-Smithsonian Center for Astrophysics, the Las Cumbres Observatory Global Telescope Network Incorporated, the National Central University of Taiwan, the Space Telescope Science Institute, the National Aeronautics and Space Administration under grant No. NNX08AR22G issued through the Planetary Science Division of the NASA Science Mission Directorate, the National Science Foundation under grant No. AST-1238877, the University of Maryland, and E{\"o}tv{\"o}s Lor{\'a}nd
University (ELTE) and the Los Alamos National Laboratory. This project used public archival data from the Dark Energy Survey (DES). Funding for the DES Projects has been provided by the U.S. Department of Energy, the U.S. National Science Foundation, the Ministry of Science and Education of Spain, the Science and Technology Facilities Council of the United Kingdom, the Higher Education Funding Council for England, the National Center for Supercomputing Applications at the University of Illinois at Urbana-Champaign, the Kavli Institute of Cosmological Physics at the University of Chicago, the Center for Cosmology and Astro-Particle Physics at the Ohio State University, the Mitchell Institute for Fundamental Physics and Astronomy at Texas A\&M University, Financiadora de Estudos e Projetos, Funda{\c c}{\~a}o Carlos Chagas Filho de Amparo {\`a} Pesquisa do Estado do Rio de Janeiro, Conselho Nacional de Desenvolvimento Cient{\'i}fico e Tecnol{\'o}gico and the Minist{\'e}rio da Ci{\^e}ncia, Tecnologia e Inova{\c c}{\~a}o, the Deutsche Forschungsgemeinschaft, and the Collaborating Institutions in the Dark Energy Survey.
The Collaborating Institutions are Argonne National Laboratory, the University of California at Santa Cruz, the University of Cambridge, Centro de Investigaciones Energ{\'e}ticas, Medioambientales y Tecnol{\'o}gicas-Madrid, the University of Chicago, University College London, the DES-Brazil Consortium, the University of Edinburgh, the Eidgen{\"o}ssische Technische Hochschule (ETH) Z{\"u}rich,  Fermi National Accelerator Laboratory, the University of Illinois at Urbana-Champaign, the Institut de Ci{\`e}ncies de l'Espai (IEEC/CSIC), the Institut de F{\'i}sica d'Altes Energies, Lawrence Berkeley National Laboratory, the Ludwig-Maximilians Universit{\"a}t M{\"u}nchen and the associated Excellence Cluster Universe, the University of Michigan, the National Optical Astronomy Observatory, the University of Nottingham, The Ohio State University, the OzDES Membership Consortium, the University of Pennsylvania, the University of Portsmouth, SLAC National Accelerator Laboratory, Stanford University, the University of Sussex, and Texas A\&M University.
This paper makes use of data based in part on observations at Cerro Tololo Inter-American Observatory, National Optical Astronomy Observatory, which is operated by the Association of Universities for Research in Astronomy (AURA) under a cooperative agreement with the National Science Foundation. 
This research uses services or data provided by the NOAO Data Lab. NOAO is operated by the Association of Universities for Research in Astronomy (AURA), Inc. under a cooperative agreement with the National Science Foundation.

\end{document}